\documentclass[aps,prl,twocolumn]{revtex4}

\begin{document}

\noindent\textbf{Comment on ``Modified Coulomb Law in a Strongly Magnetized Vacuum''}\\[-1ex]

In a recent Letter~\cite{Shabad:2007xu}, Shabad and Usov studied the electric potential of a charge placed in a strong magnetic field $B\gg B_0\equiv m^2/e$ ($m$ is the electron mass and $e$ is its charge), as modified by the vacuum polarization. According to these authors, in the limit $B\to\infty$ the modified potential becomes a Dirac delta function plus a regular background. With this potential they showed that the unboundedness from below of the nonrelativistic hydrogen spectrum is removed. In this Comment, we point out that Shabad and Usov used an incorrect calculation to arrive at their, otherwise, correct result and conclusion.

It is firmly established that the magnetic catalysis of chiral symmetry breaking~\cite{Gusynin:1994re} is a universal phenomenon. The general result states that a strong magnetic field leads to the generation of a dynamical fermion mass even at the weakest attractive interaction between fermions. The realization of this phenomenon in the chiral limit in QED has been studied in detail in the literature~\cite{Gusynin:1995gt,Gusynin:1998zq,Leung:2005yq}. In particular, in Ref.~\cite{Leung:2005yq}, we obtained an asymptotic expression for the dynamical fermion mass, reliable in the weak coupling regime and the strong field approximation.

The dynamical fermion mass in massless QED in a strong magnetic field, $m_\mathrm{dyn}$, is given by~\cite{Leung:2005yq}
\begin{equation}
m_\mathrm{dyn}\simeq\alpha\,
\exp\left(-\frac{\pi}{\alpha\log\alpha^{-1}}\right)\,\sqrt{eB},
\end{equation}
where $\alpha=e^2/4\pi\simeq 1/137$ is the fine structure constant. In a moderately strong magnetic field $B\gtrsim B_0$, the ratio $m_\mathrm{dyn}/m$ is vanishingly small hence the calculation of Ref.~\cite{Shabad:2007xu} with the nonperturbative effect of magnetic catalysis completely ignored is justified. In an infinitely strong magnetic field, however, the dynamical fermion mass $m_\mathrm{dyn}$ approaches infinity linearly with $\sqrt{eB}$. The universal nature of the magnetic catalysis of chiral symmetry breaking implies that in realistic QED with massive electrons, the electron in an infinitely strong magnetic field will acquire an effective dynamical mass that is much larger than $m$. As a result, the calculation of Ref.~\cite{Shabad:2007xu} is not correct in the limit $B\to\infty$.

Taking into consideration the nonperturbative effect of dynamical mass generation in realistic QED in a strong magnetic field, we find the vacuum polarization that is quoted in Ref.~\cite{Shabad:2007xu} to be modified by
\begin{equation}
\kappa_2(k_3^2,k_\perp^2)=-\frac{2\alpha}{\pi}\,eB\,\exp\left(-
\frac{k_\bot^2}{2eB}\right)\,T\left(\frac{k_3^2}{4m_\ast^2}\right),
\end{equation}
where the explicit dependence on $eB$ has been restored. In Eq.~(2), $m_\ast$ is the effective electron mass in a strong magnetic field, yet to be determined by a gap equation that couples to $\kappa_2$ through the full photon propagator. At this point it might appear that the only way to proceed is to first solve for $m_\ast$, which is a numerically intensive task. Fortunately, this is not the case. Since, as can be seen from Eq.~(1), the effect of magnetic catalysis increases with increase of $B$, it is conceivable that $m_\ast$ has the asymptotic behavior $m_\ast\approx m_\mathrm{dyn}\to\infty$ as $B\to\infty$. The key point is to note that while both $m_\ast$ and $B$ approach infinity in the limit $B\to\infty$, there is still a wide separation of scales in this problem, namely, $m_\ast\ll\sqrt{eB}$. Furthermore, in the limit $B\to\infty$ the ratio of these two scales, $m_\ast/\sqrt{eB}\ll 1$, is a constant independent of the magnetic field. Indeed, as shown in Ref.~\cite{Leung:2005yq}, the wide separation of scales $m_\mathrm{dyn}\ll\sqrt{eB}$ is the underlying physics behind the fact that Eq.~(1) is reliable for weak couplings and strong fields.

Based on the above analysis, the arguments of Ref.~\cite{Shabad:2007xu} that lead to an isotropic Yukawa law and, consequently, to a Dirac delta function plus a regular background in the limit $B\to\infty$ can be rendered valid, provided that we make the two corrections: (i) the electron Compton length, $m^{-1}$, is changed to the effective electron Compton length in a strong magnetic field, $m_\ast^{-1}$; (ii) the separation of scales $L_B\ll m^{-1}$ is changed to $L_B\ll m_\ast^{-1}$, where $L_B=1/\sqrt{eB}$ is the electron Larmor length. The reason that Shabad and Usov obtained the correct result and conclusion is because the separation of scales $L_B\ll m_\ast^{-1}\ll m^{-1}$ holds in the limit $B\to\infty$.

\emph{Note added in proof}. The asymptotic behavior of $m_\ast$ has been numerically verified in a recent study~\cite{Wang:2007sn}.

\vspace{2ex}
\noindent Shang-Yung Wang\\
\small{\indent Department of Physics\\
\indent Tamkang University\\
\indent Tamsui, Taipei 25137, Taiwan}\\[-1ex]


\end{document}